\documentclass[%
 reprint,
superscriptaddress,
 amsmath,amssymb,
 aps,
prb,
]{revtex4-2}

\usepackage{graphicx}
\usepackage{dcolumn}
\usepackage{bm}
\usepackage[english]{babel} 
\usepackage{hyperref}
\usepackage{physics}
\usepackage{amsmath}
\usepackage{subfigure}
\usepackage{bm}
\usepackage{epstopdf}

\begin{document}

\title{Modeling ultrafast demagnetization and spin transport: the interplay of spin-polarized electrons and thermal magnons}

\author{M. Beens}
\email[Corresponding author: ]{m.beens@tue.nl}
\affiliation{Department of Applied Physics, Eindhoven University of Technology \\ P.O. Box 513, 5600 MB Eindhoven, The Netherlands}
\author{R.A. Duine }
\affiliation{Department of Applied Physics, Eindhoven University of Technology \\ P.O. Box 513, 5600 MB Eindhoven, The Netherlands}
\affiliation{ Institute for Theoretical Physics, Utrecht University \\
Leuvenlaan 4, 3584 CE Utrecht, The Netherlands}
\author{B. Koopmans}
\affiliation{Department of Applied Physics, Eindhoven University of Technology \\ P.O. Box 513, 5600 MB Eindhoven, The Netherlands}

\date{\today}

\begin{abstract}
We theoretically investigate laser-induced spin transport in metallic magnetic heterostructures using an effective spin transport description that treats itinerant electrons and thermal magnons on an equal footing. Electron-magnon scattering is included and taken as the driving force for ultrafast demagnetization. We assume that in the low-fluence limit the magnon system remains in a quasi-equilibrium, allowing a transient nonzero magnon chemical potential. In combination with the diffusive transport equations for the itinerant electrons, the description is used to chart the full spin  dynamics within the heterostructure.  In agreement with recent experiments, we find that in case the spin-current-receiving material includes an efficient spin dissipation channel, the interfacial spin current becomes directly proportional to the temporal derivative of the magnetization. Based on an analytical calculation, we discuss that other relations between the spin current and magnetization may arise in case the spin-current-receiving material displays inefficient spin-flip scattering. Finally, we discuss the role of (interfacial) magnon transport and show that, a priori, it cannot be neglected. However, its significance strongly depends on the system parameters. 
\end{abstract}


\maketitle

\section{Introduction}

\noindent Rapidly heating magnetic heterostructures generates spin currents on ultrashort time scales \cite{malinowski2008control,Melnikov2011}. Their unique transient dynamics lead to fascinating physics in magnetic multilayers. For example, the spin current gives rise to the emission of THz electromagnetic radiation in magnetic heterostructures, resulting from the inverse spin Hall effect \cite{Kampfrath2013,Seifert2018}. Additionally, in noncollinear magnetic systems THz standing spin waves are excited by the spin-transfer torque \cite{Razdolski2017,lalieu2017absorption}. Moreover, the spin currents can play an assisting role in deterministic all-optical switching  \cite{Iihama2018,Igarashi2020,Remy2020,VanHees2020,Iihama2021}. In other words, optically induced spin currents provide a versatile tool to manipulate magnetic systems on ultrashort timescales and pave the way towards future spintronic technologies. 

Since the first experimental proof of subpicosecond demagnetization in laser-excited magnetic thin films \cite{Beaurepaire1996}, the physical origin of ultrafast spin dynamics remains a subject of heavy debate. Locally, possible mechanisms that drive ultrafast demagnetization are the direct coherent interactions between photons and spins \cite{Zhang2000,Bigot2009}, and local spin dynamics as triggered by laser heating or excitation \cite{Beaurepaire1996,Koopmans2005,Kazantseva2007,carpene2008dynamics,Krauss2009,Koopmans2010,Atxitia2011,Manchon2012,mueller2013feedback,Mueller2014,Haag2014,Nieves2014,Tveten2015,Krieger2015,Tows2015}. The latter may involve an increased rate of various spin-flip scattering processes that eventually transfer angular momentum to the lattice degrees of freedom \cite{Dornes2019,Tauchert2021}. Furthermore, nonlocal mechanisms can play a role, since spin angular momentum can be transported away from the the ferromagnetic layer via the generated spin currents. Different mechanisms have been proposed, such as  superdiffusive spin transport  \cite{Battiato2010,Battiato2012}, and the spin-dependent Seebeck effect \cite{Choi2015,Alekhin2017}.

In the last few years, multiple experimental and theoretical studies suggest that the local demagnetization and spin-current generation have the same physical origin \cite{Choi2014,Tveten2015,Kimling2017,Shin2018}. The main observation is that the rate at which spin-polarized electrons are generated is determined by the demagnetization rate \cite{Choi2014}. This can be understood as being a result of electron-magnon scattering,  which stems from the $s$-$d$ interaction that couples local magnetic moments to itinerant spins \cite{Cywinski2007,Manchon2012,Tveten2015,Gridnev2016,Beens2020}. Recent experiments support this view and show a direct proportionality between the spin current injected into a neighbouring nonmagnetic layer and the temporal derivative of the magnetization \cite{Rouzegar2021,Lichtenberg2021}. 
 
In this work, we investigate the relation between demagnetization and spin-current injection in rapidly heated magnetic heterostructures. We specifically address the role of thermal magnons and their interaction with electrons, and use a diffusive spin transport description that includes both spin-current carriers. It is assumed that electron-magnon scattering is the main driving force for ultrafast demagnetization. This scattering channel has been extensively investigated in theoretical studies \cite{Manchon2012,Tveten2015,Haag2014,carpene2008dynamics}.  Magnon transport and spin-dependent electron transport are treated on an equal footing. This is achieved by allowing the  magnon chemical potential to be nonzero \cite{cornelissen2016magnon}. The description has many similarities with the steady-state magnon transport calculations in magnetic insulators \cite{cornelissen2016magnon,rezende2014magnon,basso2016thermodynamic} and metallic heterostructures \cite{Beens2018,cheng2017interplay}. Ref.\  \cite{Beens2018} suggested that for thermally injected steady-state spin currents at metallic interfaces the magnonic contribution cannot be neglected a priori. Here, we develop this insight for the time-dependent scenario of rapidly heated magnetic heterostructures. Furthermore, we show that the interfacial spin current becomes directly proportional to the temporal derivative of the magnetization in case the receiving material is an efficient spin sink. As we demonstrate analytically, other behavior is found when the latter displays inefficient spin-flip scattering.

This article starts with an overview of the used model in Section \ref{sec:2}, specifically discussing the underlying assumptions. For a number of experimentally relevant cases, such as a Ni/Pt bilayer, we present numerical calculations for the local demagnetization and spin transport in Section \ref{sec:3a} and \ref{sec:3b}. In Section \ref{sec:3c}, we analytically derive the different relations between the interfacial spin current and the magnetization for the limiting cases of either efficient or inefficient spin dissipation in the spin-current-receiving material. Finally, we investigate the role of magnon transport and interfacial electron-magnon scattering  in more detail.

\section{Model} 
\label{sec:2}

\noindent This section gives an overview of the diffusive model we use to investigate spin transport in rapidly heated magnetic heterostructures.  Although other authors already presented the descriptions of spin-dependent electron transport
\cite{valet1993theory,Slachter2010,Kimling2017,Choi2014,Shin2018,ko2020optical} and diffusive magnon transport \cite{cornelissen2016magnon,rezende2014magnon} separately, we here discuss them in a more integrated fashion. Readers familiar with these descriptions can skip this section and move to Section \ref{sec:3}.  

Here, we start with introducing the description of the thermal magnon system. 

\subsection{Magnon density and magnon energy density} 
\label{sec:2a}

\noindent We define the magnonic system similar to Tveten et al. \cite{Tveten2015}. The standard Heisenberg model for a lattice of local spins, representing the relatively localised $3d$ electons, is expressed in terms of bosonic creation and annihilation operators  using the Holstein-Primakoff tranformation \cite{Holstein1940}. Diagonalization by the use of Fourier transformations yields the magnon dispersion relation, which is approximated as being quadratic $\epsilon_q = \epsilon_0 + A q^2$ \cite{Tveten2015}. Here, $q$ is the magnitude of the magnon wave vector, $\epsilon_0$ is the magnon gap and $A$ is the spin-wave stiffness. The corresponding density of states is then given by $D(\epsilon) = \sqrt{\epsilon-\epsilon_0}/(4\pi^2 A^{3/2})$ \cite{Bender2014,Tveten2015}. 

In contrast to Ref.\ \cite{Tveten2015}, we assume the magnon system remains internally thermalized. As we only address the low-fluence limit, we argue that after the laser pulse excites the ferromagnet the magnon distribution function remains very similar to a Bose-Einstein function.  On the ultrashort time scales that we are interested in, which can potentially be much shorter than the magnon lifetime, we should treat the magnon number as a (quasi-)conserved quantity. Then, the magnon number and total magnon energy compose two degrees of freedom. Hence, two parameters are needed to describe this system, the magnon temperature $T_m$ and the magnon chemical potential $\mu_m$. We stress that the chemical potential and temperature used here correspond to effective parameters, where effective refers to the fact that the magnon distribution function might slightly deviate from a Bose-Einstein distribution. The description is similar to Ref.\  \cite{Manchon2012}, with the extension that it allows a nonzero chemical potential. 



The magnon number density $n_d$ and magnon energy density $U_d$ are defined by the integrals \cite{Bender2014}

\begin{eqnarray}
\label{eq:nd0}
n_d &=& \int_{\epsilon_0}^\infty  d\epsilon D(\epsilon) n_\mathrm{BE} 
(\epsilon,\mu_m,T_m), \\
\label{eq:Ud0}
U_d &=& \int_{\epsilon_0}^\infty  d\epsilon (\epsilon D(\epsilon) ) n_\mathrm{BE} 
(\epsilon,\mu_m,T_m),
\end{eqnarray}

\noindent where $n_{\mathrm{BE}}(\epsilon,\mu_m,T_m)$ corresponds to the Bose-Einstein distribution 

\begin{eqnarray}
n_\mathrm{BE}(\epsilon,\mu_m,T_m) &=& 
\dfrac{1}{e^{(\epsilon-\mu_m)/(k_B T_m)}-1} .
\end{eqnarray}

\noindent Note that in Eqs.\ (\ref{eq:nd0})-(\ref{eq:Ud0}),  we extended the upper boundary of the energy integration to infinity, which is valid under the condition that the temperature remains much lower than the Curie temperature $T_m\ll T_C$. Now $n_d$ and $U_d$ can be expressed in terms of a polylogarithm \cite{Bender2012}. We assume that deviations in the magnon temperature are small compared to the ambient temperature $T_0$, i.e., $(T_m-T_0)\ll T_0$. Furthermore, we assume $\mu_m/(k_B T_0)\ll 1$ and $\epsilon_0/(k_B T_0)\ll 1$. The polylogarithm can be expressed in terms of a series expansion for the given small factors. We elimate the factors higher than linear order. Details about this approximation are given in Appendix \ref{sec:appA}.  

Following this procedure, the temporal derivative of the magnon density and the magnon energy density are expressed as

\begin{eqnarray}
\label{eq:nd}
\dfrac{\partial n_d}{\partial t} &=& C_{n,\mu} \dot{\mu}_m+C_{n,T}  \dot{T}_m ,
\\
\label{eq:Ud}
\dfrac{\partial U_d}{\partial t} &=&  C_{U,\mu} \dot{\mu}_m+
C_{U,T} \dot{T}_m .
\end{eqnarray}

\noindent The definitions of the prefactors are given in Table \ref{tab:1}. The prefactor $C_{n,\mu}$ requires special attention, since it depends on the magnon chemical potential. As explained in Appendix \ref{sec:appA}, the latter is essential to describe the correct behavior as a function of chemical potential and is a direct consequence of the bosonic nature of magnons. As the chemical potential approaches the magnon gap, the magnon density grows increasingly strong, corresponding to the divergence of $C_{n,\mu}$. For physically relevant values of the magnon gap this effect is nonnegligible. Therefore, the model includes one nonlinear term arrising from $C_{n,\mu}$. 

\subsection{Spin and energy transfer rate by electron-magnon scattering } 

\noindent Here, we give expressions for the spin transfer and energy transfer between the magnonic system and the itinerant electron system, which are driven by electron-magnon scattering. Starting from the $s$-$d$ Hamiltonian \cite{Tveten2015}, the electron-magnon scattering rate is  calculated using Fermi's golden rule \cite{Manchon2012,Bender2012,Tveten2015}. It is assumed that the itinerant electron system is instantaneously thermalized and parametrized by the  spin accumulation $\mu_s$ and electron temperature $T_e$. In the limit that the Fermi energy is the largest energy scale in the model, the angular momentum transfer rate $I_{sd}$ (in units of $\hbar$) and energy density transfer rate $U_{sd}$ can be expressed as \cite{Bender2012,Bender2014,Tveten2015}

\begin{eqnarray}
I_{sd} &=& \int_{\epsilon_0}^\infty  d\epsilon \dfrac{\Gamma(\epsilon)}{\hbar} D(\epsilon) 
(\epsilon-\mu_s) 
\\
\nonumber 
&& \qquad \times (n_{\mathrm{BE}}(\epsilon,\mu_s^F,T_e) 
-n_{\mathrm{BE}}(\epsilon,\mu_m,T_m )),
\\
U_{sd} &=& \int_{\epsilon_0}^\infty d\epsilon \dfrac{\Gamma(\epsilon)}{\hbar} (\epsilon D(\epsilon)) (\epsilon -\mu_s) 
\\
\nonumber 
&& \qquad \times (n_{\mathrm{BE}}(\epsilon,\mu_s^F,T_e) 
-n_{\mathrm{BE}}(\epsilon,\mu_m,T_m )).
\end{eqnarray}

\noindent For simplicity, the energy-dependent scattering rate coefficient $\Gamma(\epsilon)$ is assumed to be constant and  replaced by the dimensionless effective coefficient $\Gamma_0$. The constant $\Gamma_0$ can be  directly related to the effective Gilbert damping \cite{Bender2014}. 

Following the same procedure as simplifying the magnon densities, the transfer rates can be expressed as

\begin{eqnarray}
\label{eq:Isd}
I_{sd} &=& 
\dfrac{g_{n,\mu}}{\hbar} (\mu_s-\mu_m) + \dfrac{g_{n,T}}{\hbar}(T_e-T_m) ,
\\
\label{eq:Usd}
U_{sd} &=& \dfrac{g_{U,\mu}}{\hbar}  (\mu_s-\mu_m) + \dfrac{ g_{U,T}}{\hbar} (T_e-T_m).
\end{eqnarray} 

\noindent The coupling constants are summarized again  in Table \ref{tab:1}, which are all expressed in terms of the  scattering rate coefficient $\Gamma_0$, the spin-wave stiffness $A$ and the ambient temperature $T_0$. The factors $\zeta(z)$ and $\Gamma(z)$ correspond to the Riemann zeta function and Gamma function, respectively. 

\subsection{Diffusive magnon transport } 

\noindent Here, we discuss the description of diffusive magnon transport. We follow exactly the same steps as the model for diffusive magnon transport in magnetic insulators \cite{rezende2014magnon,cornelissen2016magnon}. As discussed below, applying this in ferromagnetic metals requires some extra comments. 

To treat the magnons within a local-density approximation it is needed that the characteristic length scale of the system, which in this case is the thickness of the ferromagnetic layer, is much larger than the thermal de Broglie wavelength. Up to a numerical prefactor the latter wavelength is of the order $\lambda_\mathrm{th}\sim (A/(k_B T_0))^{1/2}$ \cite{cornelissen2016magnon}. For Ni this estimate gives $\lambda_\mathrm{th}\sim 0.4\mbox{ nm}$ at room temperature, using the numerical values listed in Table \ref{tab:2}. Secondly, to be able to describe the transport as diffusive the magnon mean free path $\lambda_\mathrm{mfp}\sim (A k_B T_0)^{1/2} \tau_{\mathrm{tr},m} /\hbar$ should be much smaller than the thickness of the ferromagnetic system. The magnon momentum relaxation time $\tau_{\mathrm{tr},m}$ is discussed below. For Ni we estimate that the mean free path is of the order $\lambda_\mathrm{mfp}\sim 1.5 \mbox{ nm}$. Despite that these requirements are only weakly satisfied for an ultrathin ferromagnetic layer, we assume that the qualitative behavior is predicted correctly by the diffusive magnon transport description. 

Within these limits the magnon current density and the magnon heat current density can be expressed as \cite{cornelissen2016magnon}

\begin{eqnarray}
\label{eq:jm}
j_m &=& -\dfrac{\sigma_m}{e^2} \dfrac{\partial \mu_m}{\partial x} 
-\dfrac{L}{T_0} \dfrac{\partial T_m}{\partial x} ,
\\
\label{eq:jQm}
j_{Q,m} &=& 
-L\dfrac{\partial \mu_m}{\partial x} 
-\kappa_m \dfrac{\partial T_m}{\partial x} ,
\end{eqnarray}

\noindent  where $\sigma_m$ is the magnon conductivity, $L$ is the spin Seebeck coefficient \cite{rezende2014magnon,Uchida2010}, and $\kappa_m$ is the magnon heat conductivity. The transport coefficients are given in Table \ref{tab:1}. To a good approximation, all transport coefficients are linear in the magnon transport time scale $\tau_{\mathrm{tr},m}$, which corresponds to the magnon momentum relaxation time. This time scale is at least as short as the electron-magnon scattering time, which is naturally related to the observed demagnetization time scale. Hence, the latter is an upper bound for $\tau_{\mathrm{tr},m}$. In the remainder of this article we assume that the time scale $\tau_{\mathrm{tr},m}$ is of the same order of magnitude as the demagnetization time. For instance, we use $\tau_{\mathrm{tr},m}=0.1\mbox{ ps}$, corresponding to the typical order of magnitude of the demagnetization time in ferromagnetic transition metals.    

To clarify the notation we give the continuity equation for the magnon density and magnon energy density 

\begin{eqnarray}
\label{eq:nd2}
\dfrac{\partial n_d}{\partial t} 
+\dfrac{\partial j_m}{\partial x} 
&=& I_{sd} ,
\\
\label{eq:Ud2}
\dfrac{\partial U_d}{\partial t} 
+\dfrac{\partial j_{Q,m} }{\partial x} 
&=& U_{sd} .
\end{eqnarray}

\noindent Filling in Eqs.\ (\ref{eq:nd})-(\ref{eq:Ud}), Eqs.\ (\ref{eq:Isd})-(\ref{eq:Usd}), and  Eqs.\ (\ref{eq:jm})-(\ref{eq:jQm}), gives the full expressions that are used in the calculations presented in the later sections of this article. Now we move on to the electronic system.

\begin{table}
\centering
\caption{\label{tab:1} Model coefficients expressed in terms of the magnon transport time scale $\tau_{\mathrm{tr},m}$, spin-wave stiffness $A$ and the bulk electron-magnon scattering rate coefficient $\Gamma_0$  \cite{cornelissen2016magnon}. The interfacial scattering rate coefficients can be found by the substitution $\Gamma_0\rightarrow g_{\uparrow\downarrow}/(\pi s)$.  }
\begin{tabular}[t]{lc}
\hline  \\ [-2.0ex]
Symbol &   Expression \\
\hline  \\ [-1.5ex] \vspace{1mm}
$C_{n,\mu}$ &   $\dfrac{(k_B T_0)^{1/2}\Gamma(3/2)}{4\pi^2 A^{3/2} }
\Big(\Gamma(1/2) \Big(\dfrac{\epsilon_0-\mu_m}{k_B T_0}\Big)^{-1/2}
+\zeta(1/2)   
\Big)
$\\  \vspace{1mm}
$C_{n,T}$ &  $\dfrac{(k_B T_0)^{1/2}\Gamma(3/2)}{4\pi^2 A^{3/2} }
(3/2)\zeta(3/2) k_B$   \\  \vspace{1mm}
$C_{U,\mu}$  & $\dfrac{(k_B T_0)^{3/2}\Gamma(5/2)}{4\pi^2 A^{3/2} } \zeta(3/2) $ \\  \vspace{1mm}
$C_{U,T}$ & $\dfrac{(k_B T_0)^{3/2}\Gamma(5/2)}{4\pi^2 A^{3/2} }(5/2) \zeta(5/2) k_B $  \\ \vspace{1mm}
$\sigma_m $ &  
$\dfrac{e^2\tau_{\mathrm{tr},m}(k_B T_0)^{3/2} \Gamma(3/2)\zeta(3/2)}{2\pi^2 \hbar^2 A^{1/2}} $ \\  \vspace{1mm}
$L $ &  $\dfrac{\tau_{\mathrm{tr},m}(k_B T_0)^{5/2}(5/2) \Gamma(5/2)\zeta(5/2)}{3 \pi^2 \hbar^2 A^{1/2}}  $ \\  \vspace{1mm}
$\kappa_m $ & $\dfrac{\tau_{\mathrm{tr},m}(k_B T_0)^{5/2}(7/2) \Gamma(7/2)\zeta(7/2)}{3 \pi^2 \hbar^2 A^{1/2}} k_B $ \\  \vspace{1mm}
$g_{n,\mu}$ &  $ \dfrac{\Gamma_0 (k_B T_0)^{3/2}\Gamma(5/2)\zeta(3/2) }{4\pi^2 A^{3/2} } $  \\  \vspace{1mm} 
$g_{n,T}$ &   $\dfrac{\Gamma_0(k_B T_0)^{3/2}\Gamma(5/2)(5/2)\zeta(5/2) }{4\pi^2 A^{3/2} } k_B $  \\  \vspace{1mm}
$g_{U,\mu}$ & $\dfrac{\Gamma_0(k_B T_0)^{5/2}\Gamma(7/2)\zeta(5/2) }{4\pi^2 A^{3/2} } $  \\  \vspace{1mm} 
$g_{U,T}$ &   $\dfrac{\Gamma_0(k_B T_0)^{5/2}\Gamma(7/2)(7/2)\zeta(7/2) }{4\pi^2 A^{3/2} } k_B $  \\  
\hline
\end{tabular}
\end{table}

\subsection{The continuity equations for the electronic system}

\noindent We assume that the out-of-equilibrium spin density $\delta n_s$ in the itinerant electron system can be parametrized by $\delta n_s = \tilde{\nu}_\mathrm{F} \mu_s^\mathrm{F}$, where $\tilde{\nu}_\mathrm{F} =2 \nu_\uparrow \nu_\downarrow/(\nu_\uparrow+\nu_\downarrow) $ is the spin-averaged density of states evaluated at the Fermi energy \cite{Note1}. Expressed in terms of the spin accumulation, the continuity equations for the spin and energy in the ferromagnetic layer are given by \cite{Kimling2017,Tveten2015}

\begin{eqnarray}
\label{eq:ns}
\tilde{\nu}_\mathrm{F} \dfrac{\partial \mu_s^\mathrm{F} }{\partial t} 
+\dfrac{\partial j_{s,e}^\mathrm{F} }{\partial x} ,
 &=& 
-\dfrac{\tilde{\nu}_\mathrm{F} \mu_s^\mathrm{F} }{\tau_{s,\mathrm{F}} } 
-2I_{sd} ,
\\
\label{eq:Us}
C_{e} \dfrac{\partial T_e^\mathrm{F}}{\partial t}
+\dfrac{\partial j_{Q,e}^\mathrm{F} }{\partial x} 
&=& g_{ep}(T_p^\mathrm{F}-T_e^\mathrm{F})-U_{sd} + P(t,x) .
\end{eqnarray}

\noindent The spin current $j_{s,e}^\mathrm{F}$ and electronic heat current $j_{Q,e}^\mathrm{F}$ are given below. The term proportional to $\tau_{s,\mathrm{F}}^{-1}$, which is introduced phenomenologically \cite{Tveten2015}, represents the additional spin-flip scattering processes. The latter includes Elliott-Yafet spin-flip scattering processes and is the main spin dissipation channel for the combined electronic and magnonic system \cite{Tveten2015}. $C_e$ corresponds to the electron heat capacity,  $g_{ep}$ corresponds to the electron-phonon coupling constant and $T_p^\mathrm{F} $ corresponds to the phonon temperature. The function $P(t,x)$ represents the laser-excitation profile, which will be further specified when the calculations are presented. 

Imposing that there is no charge transport, the electronic spin current $j_{s,e}^\mathrm{F}$ and heat current $j_{Q,e}^\mathrm{F}$ can be expressed as \cite{Slachter2010,Kimling2017}

\begin{eqnarray}
j_{s,e}^\mathrm{F} &=& -\dfrac{\tilde{\sigma} }{e^2} 
\dfrac{\partial \mu_s^\mathrm{F}}{\partial x}  -\dfrac{\tilde{\sigma}}{e^2} 
S_s \dfrac{\partial T_e^\mathrm{F}}{\partial x}  ,
\\
j_{Q,e}^\mathrm{F} &=& -\dfrac{\tilde{\sigma}}{2e^2} 
\Pi_s
\dfrac{\partial  \mu_s^\mathrm{F} }{\partial x} 
-\kappa_{e} \dfrac{\partial T_e^\mathrm{F}}{\partial x} ,
\end{eqnarray}

\noindent where $\tilde{\sigma}=2\sigma_\uparrow \sigma_\downarrow/(\sigma_\uparrow +\sigma_\downarrow)$ is the spin-averaged electrical conductivity, $S_s$ is the spin-dependent Seebeck coefficient \cite{Slachter2010}, $\Pi_s$ is the spin-dependent Peltier coefficient ($\Pi_s=T_0 S_s$), and $\kappa_{e} $ is the electronic heat conductivity. The expressions for the dynamics of the electronic system within the nonmagnetic layer can be found by replacing all indices $\mathrm{F}\rightarrow\mathrm{N}$ and removing the spin-dependent quantities (including $I_{sd}$ and $U_{sd}$).

Finally, for the phonon system we take a highly simplified approach. For convenience, phonon heat transport is not included. Furthermore, in the description for the local phonon temperature a heat sink is included that dissipates energy out of the phonon system within a time scale of $20\mbox{ ps}$. The latter is introduced to make sure the system relaxes to its initial temperature on a reasonable time scale. We stress that the exact description of the phonon system does not play a direct role in the discussions presented in this work.

\subsection{Boundary conditions and system specifications}
\label{sec:2e}

\noindent Finally, we have to specify the boundary conditions. We define the ferromagnetic layer on the domain $ x\in [-d_\mathrm{F},0]$, where $d_\mathrm{F}$ is the thickness of the ferromagnetic layer. At the left end of the system we impose insulating boundary conditions, setting all currents to zero.

\begin{eqnarray}
j_m(-d_\mathrm{F})=j_{Q,m}(-d_\mathrm{F}) &=& 0.
\\
j_{s,e}^\mathrm{F}(-d_\mathrm{F} )=j_{Q,e}^\mathrm{F}(-d_\mathrm{F} ) &=& 0.
\end{eqnarray}

\noindent Secondly, at the interface, which is positioned at $x=0$, the total spin current and total heat current should be continuous.

\begin{eqnarray}
\label{eq:jstot0}
j_{s,e}^\mathrm{F}(0)+ 2 j_m(0) &=& j_{s,e}^\mathrm{N}(0) ,
\\
j_{Q,e}^\mathrm{F}(0)+j_{Q,m}(0) &=& j_{Q,e}^\mathrm{N}(0) ,
\end{eqnarray}

\noindent where the superscript $\mathrm{N}$ indicates the quantities in the nonmagnetic layer. The factor $2$ arises from the fact that a magnon carries twice as much spin angular momentum as an electron. We write the interfacial electronic spin current and heat current as 

\begin{eqnarray}
j^\mathrm{F}_{s,e}(0)&=& \dfrac{g_1}{\hbar} 
(\mu_s^\mathrm{F}-\mu_s^\mathrm{N} ) 
+\dfrac{g_1}{\hbar}  S_s^i (T_e^\mathrm{F}-T_e^\mathrm{N}) ,
\\
j^\mathrm{F}_{Q,e}(0)&=& \dfrac{g_1}{\hbar} \dfrac{T_0 S_s^i}{2} 
(\mu_s^\mathrm{F}-\mu_s^\mathrm{N} ) 
+\kappa_e^i (T_e^\mathrm{F}-T_e^\mathrm{N}) ,
\end{eqnarray}

\noindent where the prefactor $g_1$ is determined by the interfacial electrical conductance \cite{Shin2018} and all variables are evaluated at $x=0$. The interfacial electronic heat conductivity is given by $\kappa_e^i $. The factor $S_s^i$ corresponds to the spin-dependent Seebeck coefficient of the interface. 

The interfacial magnon current and magnon heat current are determined by the interfacial electron-magnon scattering rate \cite{Tveten2015,Beens2018}. The linearized expressions for the scattering rate can be found by replacing $\Gamma_0\rightarrow g_{\uparrow\downarrow}/(\pi s)$  in Eqs.\ (\ref{eq:Isd})-(\ref{eq:Usd}) \cite{Bender2012,Bender2014,Tveten2015}, where $g_{\uparrow\downarrow}$ is the real part of the spin-mixing conductance and $s$ is the saturation spin density. In other words, the interfacial magnon current $j_m(0)$ and magnon heat current $j_{Q,m}(0)$ are expressed as 

\begin{eqnarray}
\label{eq:jm0}
j_m(0) &=& 
\dfrac{g_{n,\mu}^i}{\hbar} (\mu_m-\mu_{s}^\mathrm{N}) +\dfrac{ g_{n,T}^i}{\hbar} (T_m-T_{e}^\mathrm{N}) ,
\\
j_{Q,m}(0) &=& \dfrac{g_{U,\mu}^i}{\hbar} (\mu_m-\mu_{s}^\mathrm{N}) 
+ \dfrac{g_{U,T}^i}{\hbar} (T_m-T_{e}^\mathrm{N}).
\end{eqnarray} 

\noindent The second term in Eq.\ (\ref{eq:jm0}), proportional to $g^i_{n,T}$, corresponds to the interfacial spin Seebeck effect \cite{Xiao2010}. 

Finally, the nonmagnetic layer is defined on the domain $x\in[0,d_\mathrm{N}]$, where $d_\mathrm{N}$ is the thickness of the nonmagnetic layer. At the outer interface $x=d_\mathrm{N}$ we impose the boundary conditions

\begin{eqnarray}
\label{eq:jsdN}
j_{s,e}^\mathrm{N}(d_\mathrm{N}) &=& \dfrac{g_2}{\hbar} \mu_{s}^\mathrm{N}(d_\mathrm{N})  ,
\\
j_{Q,e}^\mathrm{N}(d_\mathrm{N}) &=& 0.
\end{eqnarray}

\noindent For convenience, we assume that this interface is a heat insulator. In contrast, we allow the interface to be permeable for spins. The latter is parametrized by the constant $g_2$. In case $g_2\neq 0 $, spins are allowed to leak out of the bilayer. It is assumed  that the interface is connected to an ideal spin sink, which corresponds to a vanishing $\mu_s$ for $x > d_\mathrm{N}$ and yields Eq.\ (\ref{eq:jsdN}). The latter could for example be realized by a secondary magnetic layer that is perpendicularly oriented to the other magnetic layer, as is the case in noncollinear magnetic heterostructures \cite{Lichtenberg2021}.

In the following, we will start with discussing the situation where $g_2=0$, corresponding to completely insulating boundary conditions. Later, we will investigate the situation $g_2\neq 0$ in more detail.

\begin{figure}[t!]
\includegraphics[scale=0.95]{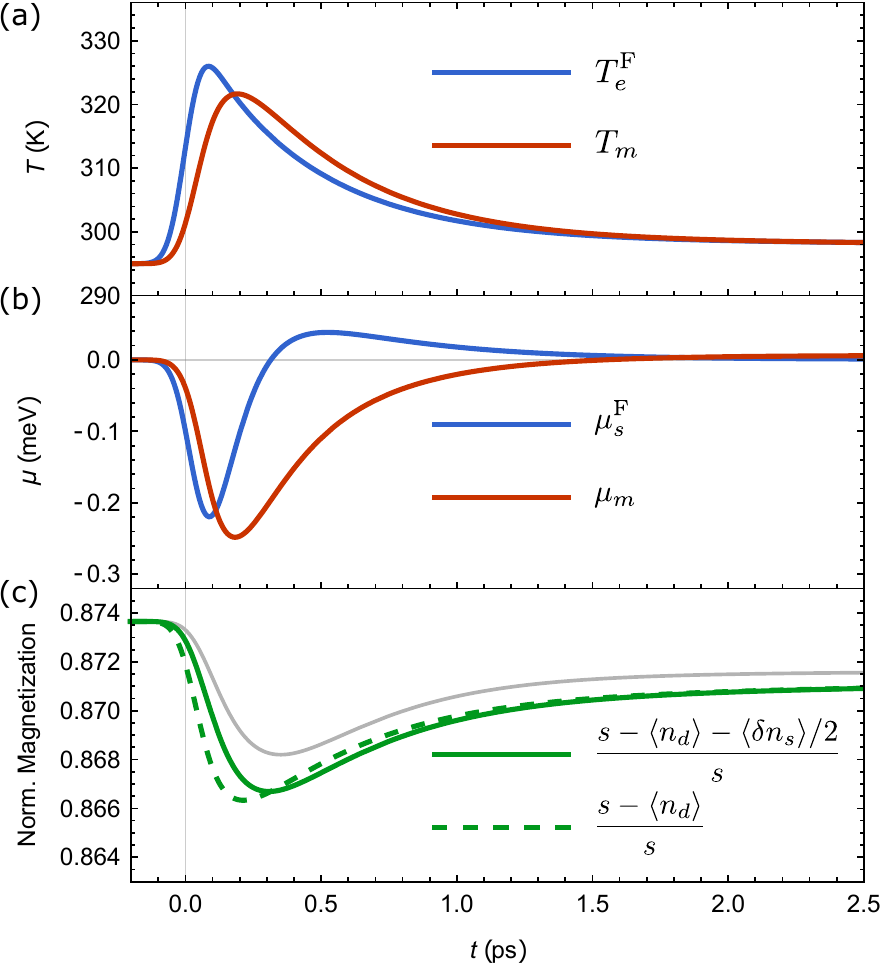}
\caption{\label{fig:1} Laser-induced dynamics of the temperatures, chemical potentials and magnetization within the Ni layer of a Ni($5\mbox{ nm}$)/Pt($3\mbox{ nm}$) bilayer. The quantities are plotted as a function of $t$ after laser-pulse excitation at $t=0$, and are spatially averaged over the ferromagnetic (Ni) layer. (a) The electron temperature (blue) and magnon temperature (red). (b) The spin accumulation (blue)  and magnon chemical potential (red). (c) The normalized magnetization. The solid green line indicates the changes of the total spin density. The dashed green line indicates the case that only changes in magnon density are taken into account. The thin gray line represents the magnetization in case of an isolated Ni layer, in the absence of a neighboring Pt layer. 
} 
\end{figure}

\section{Results} 
\label{sec:3}

\noindent In this section we present the numerical solutions to the equations discussed in the previous sections. Specifically, Eqs.\ (\ref{eq:nd2})-(\ref{eq:Ud2}) are solved for the magnonic system in the ferromagnetic layer, and Eqs.\ (\ref{eq:ns})-(\ref{eq:Us}) are solved for the electronic system throughout the complete heterostructure. Furthermore, the boundary conditions as discussed in Section \ref{sec:2e} are imposed. First, we investigate the dynamics of the local thermodynamical parameters: temperatures, chemical potentials and the magnetization.


\subsection{Temperature, chemical potential and magnetization dynamics} 
\label{sec:3a}  

\noindent We start with calculating the laser-induced response of a Ni($5\mbox{ nm}$)/Pt($3\mbox{ nm}$) bilayer with insulating boundary conditions at the outer interfaces. Specifically, we first investigate the dynamics of the local thermodynamical parameters within the Ni layer. To model laser heating we assume that the spatial and temporal profile of the laser pulse can be approximated by 

\begin{equation}
P(t,x) = \dfrac{P_0}{\sigma \sqrt{\pi}} \exp [-\dfrac{x+d_\mathrm{F}}{\tilde{\lambda}}] \exp [-\dfrac{t^2}{\sigma^2}],
\end{equation}

\noindent  where $P_0$ is the absorbed laser pulse energy density and $\sigma$ determines the pulse duration, which is set to  $70\mbox{ fs}$. The laser pulse penetration depth is given by $\tilde{\lambda}$ and set to a typical value of $\tilde{\lambda}=15\mbox{ nm} $. For simplicity, we assume that the laser pulse absorption in the Ni and Pt layer is equally efficient and we use $P_0=0.15\times 10^8\mbox{ Jm}^{-3}$. All other system parameters are given in Table \ref{tab:2} and \ref{tab:3}.

Figures \ref{fig:1}(a)-(c) show the response of the magnetic bilayer to laser heating. All plotted variables are spatially averaged over the range of the ferromagnetic (Ni) layer. Figure \ref{fig:1}(a) shows the rapid increase of the electron temperature $T_e^\mathrm{F}$ and the response of the magnon temperature $T_{m}$ driven by electron-magnon scattering. This transient behavior of the magnon temperature yields a rapid increase in the magnon density. Figure \ref{fig:1}(b) displays the laser-induced  dynamics of the the spin accumulation (blue) and magnon chemical potential (red). The spin accumulation shows the typical bipolar behavior, in analogy with previous calculations and experimental observations of the generated spin-polarized electrons \cite{Choi2014,Beens2020,Shin2018,Kimling2017}. The magnon chemical potential shows different behavior, it can be shown that this is related to that the equilibration of the chemical potentials plays a minor role and the magnon chemical potential opposes the dynamics of the magnon temperature. 


Finally, Fig.\ \ref{fig:1}(c) shows the normalized magnetization as a function of time. The magnetization requires special attention. In this work, it is assumed that the magnetic signal measured in the experiments is determined by the total spin density. The magnetization is defined as 

\begin{equation}
m = \dfrac{s-\langle n_d \rangle-\langle \delta n_s
\rangle /2}{s} = \dfrac{s-\langle n_\mathrm{tot}\rangle}{s}  , 
\end{equation}

\noindent which is normalized with respect to the saturation spin density  $s=S/a^3$, where $S$ is the spin per atom (in units of $\hbar$) and $a$ the lattice constant. The bracket notation indicates spatial averaging over the ferromagnetic layer. The solid green curve in Fig.\ \ref{fig:1}(c) shows the typical ultrafast demagnetization behavior and critically depends on the spin-flip scattering rate $\tau_{s,\mathrm{F}}$.  Since electron-magnon scattering conserves the total spin angular momentum, Elliott-Yafet spin-flip processes originating from spin-orbit coupling enable the demagnetization of the combined spin system \cite{Tveten2015,Haag2014,Fahnle2015,illg2013ultrafast}. In the end, spin is efficiently transferred to the lattice, as was recently demonstrated experimentally \cite{Dornes2019,Tauchert2021}. We stress that this interpretation of the magnetization remains a point of discussion and its relation to the magnetic signal in the experiments strongly depends on the probing method. Therefore, we have plotted the dynamics of the magnon density $\langle n_d\rangle$ separately, normalized with respect to the saturation spin density $s$, indicated by the dashed green line in Fig.\ \ref{fig:1}(c). As will be discussed in the next section, the used interpretation of the magnetization, as being determined by the sum of the magnon density and the itinerant electron spin density, is strongly supported by the investigation of the relation between the interfacial spin current and the demagnetization rate. 

Finally, the thin gray line in Fig.\ \ref{fig:1}(c) represents the calculation of the magnetization in case the Pt layer is absent and spin can not be transported out of the Ni layer. The latter emphasizes that, although interfacial spin transfer yields a significant increase of the demagnetization rate (as compared to the solid green curve), the demagnetization is primarily driven by local spin dissipation. 

\begin{figure}[t!]
\includegraphics[scale=0.95]{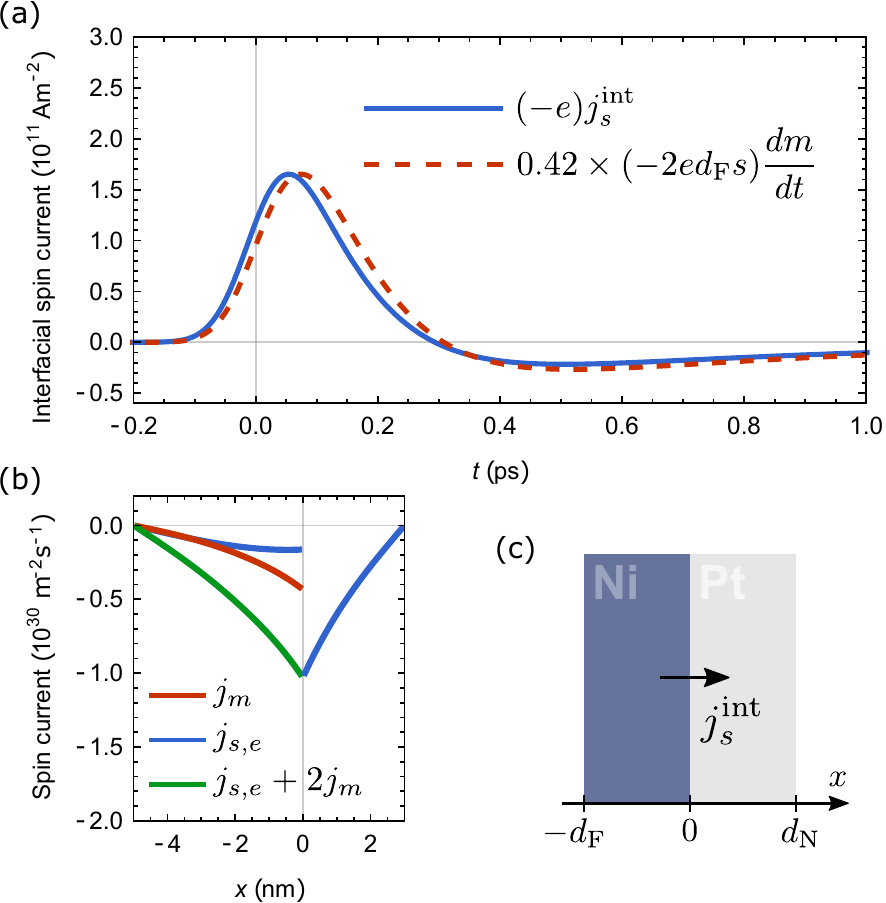}
\caption{\label{fig:2} Laser-induced spin transport in a Ni($5\mbox{ nm}$)/Pt($3\mbox{ nm}$) bilayer with insulating boundary conditions. (a) The interfacial spin current (blue) as a function of time $t$ after laser-pulse excitation at $t=0$. The dashed red line indicates the temporal derivative of the magnetization, scaled by a prefactor that is fitted to the amplitude of the spin current. (b) Distinct spin current contributions as a function of spatial coordinate $x$, evaluated at $t=0.05\mbox{ ps} $. Blue indicates the electronic contribution, red the magnonic contribution and green the total. (c) Schematic overview of the system.
} 
\end{figure}

\subsection{Spin transport in magnetic heterostructures }
\label{sec:3b} 

\noindent In this section, we calculate the spin current that arises from laser exciting the magnetic heterostructure corresponding to the results of Fig.\ \ref{fig:1}. As given by Eq.\ (\ref{eq:jstot0}), we define the total interfacial spin current at $x=0$ as $j_s^\mathrm{int}=j_{s,e}^\mathrm{F}(0)+2j_m(0)$, where $j_{s,e}^\mathrm{F}(0)$ is the spin current carried by the conduction electrons and $j_m(0)$ is the interfacial magnon current. We again  focus on a Ni($5\mbox{ nm}$)/Pt($3\mbox{ nm}$) bilayer with insulating boundary conditions, as schematically depicted in Fig.\ \ref{fig:2}(c). 

The blue line in Fig.\ \ref{fig:2}(a)  shows the results from calculating $j_s^\mathrm{int}$ by numerically solving the set of equations as presented in Section \ref{sec:2}. The material parameters and description of the laser pulse are identical to the previous section. The result clearly shows the bipolar behavior of the interfacial spin current, yielding a transient oscilation within the THz regime. The red dashed curve indicates the temporal derivative of the magnetization, scaled by a prefactor that is fitted to the amplitude of the spin current. The comparison indicates a qualitative agreement with the experiments \cite{Rouzegar2021,Lichtenberg2021}, as a close relation is expected between the spin current injected into the nonmagnetic layer and the temporal derivative of magnetization. The visible phase shift is interesting in itself, but smaller than the temporal resolution of $40\mbox{ fs}$ in the experiment in Ref.\  \cite{Rouzegar2021}.

Figure \ref{fig:2}(b) shows the different contributions to the spin current as a function of position $x$, calculated at $t=0.05\mbox{ ps}$. The figure suggests that for the used parameters magnon transport and spin-polarized electron transport comparably contribute to the total spin current within the bulk of the ferromagnet. One should keep in mind that their ratio strongly depends on the specific time instance and system parameters. The spin transport by electrons is mainly driven by bulk electron-magnon scattering, which generates negatively polarized spins that are transferred towards the receiving layer via spin diffusion. The negative magnon current in Fig.\ \ref{fig:2}(b) indicates thermal magnons being created at the interface by  electron-magnon scattering. Consequently, a flow of magnons towards the negative $x$ direction is generated. The magnon current $j_m(0)$ is mainly determined by the temperature difference $T_m-T_e^\mathrm{N}$ at the interface, which corresponds to the interfacial spin Seebeck effect \cite{Xiao2010}. In contrast to our work, the latter is typically neglected in the models for spin transport in metallic magnetic heterostructures \cite{Kimling2017,Shin2018}.

In the following section we investigate the relation between the interfacial spin current and the magnetization analytically, and specifically address the role of (interfacial) magnon transport.

\subsection{Relation between the interfacial spin current and demagnetization}
\label{sec:3c} 

\noindent In this section we analytically investigate the relation between the interfacial spin current and the demagnetization. 

Integrating Eqs.\ (\ref{eq:nd2}) and (\ref{eq:ns}) over the thickness of the ferromagnetic layer and adding up the results yields an expression for the interfacial spin current 

\begin{equation}
\label{eq:jsi1}
j_s^\mathrm{int}(t) 
= -2 d_\mathrm{F} \dfrac{d \langle n_\mathrm{tot} \rangle}{d t} 
-d_\mathrm{F} \dfrac{\langle \delta n_s\rangle }{\tau_{s,\mathrm{F}} } ,
\end{equation}

\noindent where the brackets indicate spatial averaging over the ferromagnetic layer. Equation (\ref{eq:jsi1}) simply follows from spin angular momentum conservation, as the total spin density $\langle n_\mathrm{tot}\rangle $ can only be changed by either spin transport or local spin dissipation. In the limit where the latter is absent $\tau_{s,\mathrm{F}}\rightarrow \infty$, spin transport and demagnetization couple trivially. For the systems of interest, where we have a subpicosecond $\tau_{s,\mathrm{F}}$, a more cumbersome calculation is required to eliminate the local spin dissipation term from Eq.\ (\ref{eq:jsi1}). 

In order to do this, we solve the spin diffusion equation for the full heterostructure. In the frequency domain, we write

\begin{eqnarray}
\dfrac{\partial^2 \mu_s^\mathrm{F}(\omega ,x)}{\partial x^2}  
&=&
\kappa_\mathrm{F}(\omega)^2  \mu_{s}^\mathrm{F}(\omega,x)
+\dfrac{2\tau_{s,\mathrm{F}} I_{sd}(\omega,x )  }{\tilde{\nu}_\mathrm{F} l_{s,\mathrm{F}}^2},
\\
\dfrac{\partial^2 \mu_s^\mathrm{N}(\omega ,x)}{\partial x^2}  
&=&
\kappa_\mathrm{N}(\omega)^2  \mu_{s}^\mathrm{N}(\omega,x),
\end{eqnarray}

\noindent where we use the parameter $\kappa_\mathrm{F}(\omega)= l_{s,\mathrm{F}}^{-1} \sqrt{i\omega\tau_{s,\mathrm{F}} +1}  $ for the ferromagnetic layer and $\kappa_\mathrm{N}(\omega) = l_{s,\mathrm{N}}^{-1}  \sqrt{i\omega\tau_{s,\mathrm{N}} +1} $ for the nonmagnetic layer \cite{tserkovnyak2005nonlocal}. $l_{s,\mathrm{F}}$ and $l_{s,\mathrm{N}}$ correspond to the spin diffusion length of the ferromagnetic and nonmagnetic layer respectively. The boundary conditions are identical to Section \ref{sec:2e}, but now expressed in the frequency domain. For convenience, we neglect the spin-dependent Seebeck effect ($S_s$ and $S_s^i$) in this analytical calculation. The goal is to express the Fourier transform of the interfacial spin current

\begin{equation}
j_s^\mathrm{int}(\omega) =
\dfrac{g_1}{\hbar} 
(\mu_{s,\mathrm{F}}(\omega,0)-\mu_{s,\mathrm{N}}(\omega,0)) + 2 j_m(\omega,0) ,
\end{equation}

\noindent in terms of the electron-magnon scattering rates, specifically, the bulk contribution $I_{sd}(\omega,x)$ and interfacial contribution $j_m(\omega,0)$. The resulting expression is given by 

\begin{equation}
\label{eq:jsi2}
j_s^\mathrm{int} (\omega)  = 2 A(\omega) j_m(\omega,0) -2 d_\mathrm{F} B(\omega) \mathcal{I}_{sd}(\omega) ,
\end{equation}

\noindent where $\mathcal{I}_{sd}(\omega)$ is given by 

\begin{equation}
\label{eq:source}
\mathcal{I}_{sd}(\omega)  =
\int_{-d_\mathrm{F}}^0 dx'  
\dfrac{\kappa_\mathrm{F}(\omega) \cosh[(d_\mathrm{F}+x')\kappa_\mathrm{F}(\omega)]}{\sinh[d_\mathrm{F}\kappa_\mathrm{F}(\omega) ]}
I_{sd} (\omega,x') .
\end{equation}

\noindent Furthermore, the function $A(\omega)$ is given by

\begin{equation}
\label{eq:A}
A(\omega) =  \dfrac{1 + \dfrac{ \hbar  \tilde{\nu}_\mathrm{F} d_\mathrm{F}}{g_1 \tau_{s,\mathrm{F}}}\bigg(\dfrac{l_{s,\mathrm{F}}^2\kappa_\mathrm{F}(\omega)}{d_\mathrm{F}}\bigg) \tanh[\kappa_\mathrm{F}(\omega) d_\mathrm{F} ]}{1+
\dfrac{\hbar \tilde{\nu}_\mathrm{F} d_\mathrm{F} }{g_1\tau_{s,\mathrm{F}}} 
\bigg(\dfrac{l_{s,\mathrm{F}}^2\kappa_\mathrm{F}(\omega) }{d_\mathrm{F}}\bigg) \tanh[\kappa_\mathrm{F}(\omega) d_\mathrm{F} ]
G_\mathrm{N}(\omega)} ,
\end{equation}

\noindent which can deviate from one (compared to Eq.\ (\ref{eq:jsi2})), indicating that the spin current driven by electron-magnon scattering at the interface is modified by spins flowing back into the ferromagnetic layer. Secondly, the bulk contribution depends on the function $B(\omega)$

\begin{equation}
\label{eq:B}
B(\omega ) = \dfrac{  \bigg(\dfrac{1}{d_\mathrm{F} \kappa_\mathrm{F}(\omega) }\bigg) \tanh[\kappa_\mathrm{F}(\omega) d_\mathrm{F} ]}{1+
\dfrac{\hbar \tilde{\nu}_\mathrm{F} d_\mathrm{F} }{g_1\tau_{s,\mathrm{F}} } 
\bigg(\dfrac{l_{s,\mathrm{F}}^2\kappa_\mathrm{F}(\omega) }{d_\mathrm{F}}\bigg) \tanh[\kappa_\mathrm{F}(\omega)  d_\mathrm{F} ]
G_\mathrm{N}(\omega) } .
\end{equation}

\noindent The function $G_\mathrm{N}(\omega)$ includes all the parameters that describe the properties of the nonmagnetic layer and the interfaces

\begin{equation}
G_\mathrm{N}(\omega) = 1+\dfrac{g_1+\dfrac{g_1 g_2 \tau_{s,\mathrm{N}}}{\hbar \tilde{\nu}_\mathrm{N} l_{s,\mathrm{N}}^2\kappa_\mathrm{N}(\omega) } \tanh[\kappa_\mathrm{N}(\omega) d_\mathrm{N}] }{g_2+
\dfrac{\hbar \tilde{\nu}_\mathrm{N} l_{s,\mathrm{N}}^2\kappa_\mathrm{N}(\omega) }{\tau_{s,\mathrm{N}} }\tanh[\kappa_\mathrm{N}(\omega) d_\mathrm{N} ] }.
\end{equation} 

\noindent What remains is simplifying Eq.\ (\ref{eq:jsi2}) and expressing it in terms of the total spin density $\langle n_\mathrm{tot}(\omega) \rangle$ and thereby the normalized magnetization $m$. For convenience, we first focus on the situation that interfacial electron-magnon scattering is absent.

\begin{figure}[t!]
\includegraphics[scale=0.95]{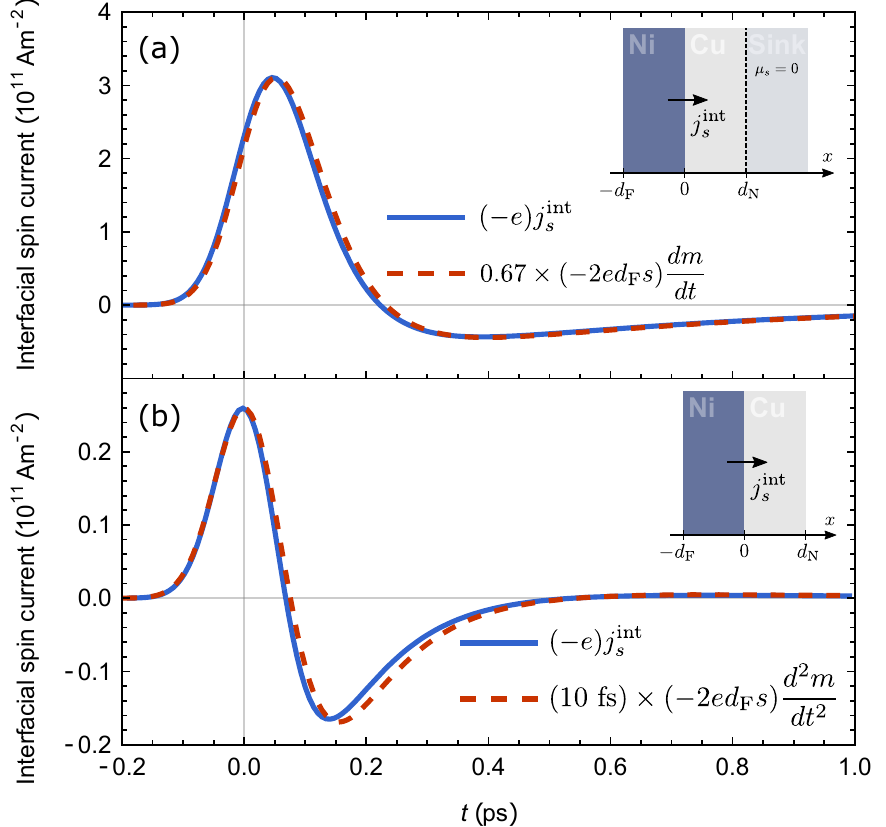}
\caption{\label{fig:3} Laser-induced spin transport in a Ni($3.4\mbox{ nm}$)/Cu($2.5\mbox{ nm}$) bilayer for two types of boundary conditions at $x=d_\mathrm{N}$. (a) The interfacial spin current (blue) as a function of time $t$ after laser-pulse excitation at $t=0$, for an interface at $x=d_\mathrm{N}$ that is permeable for spins ($g_2\neq 0$) and connected to an ideal spin sink. The dashed red line indicates the temporal derivative of the magnetization, scaled by a prefactor that is fitted to the amplitude of the spin current. (b) The interfacial spin current (blue) for insulating boundary conditions ($g_2=0$). The dashed red line indicates the second derivative of the magnetization, scaled to have the same amplitude as the spin current. 
} 
\end{figure}

\subsubsection{Bulk electron-magnon scattering}

\noindent Here, we set $j_m(\omega,0)\rightarrow 0$. Hence, the interfacial spin current is given by the second term in Eq.\ (\ref{eq:jsi2}). As is discussed in Appendix \ref{sec:appB} , a relevant approximation is that the function $\mathcal{I}_{sd}(\omega)$ closely resembles the spatial average $\langle I_{sd} (\omega) \rangle $. To eliminate $\langle I_{sd}(\omega)\rangle $ from Eq.\ (\ref{eq:jsi2}), we make use of the continuity equations for $n_d$ and $\delta n_s$ spatially averaged over the ferromagnetic layer. In the absence of interfacial electron-magnon scattering this yields

\begin{eqnarray}
\label{eq:ndomega}
i\omega \langle n_d(\omega)  \rangle 
&=& \langle I_{sd}(\omega )\rangle,
\\
\label{eq:nsomega} 
i\omega \langle \delta n_s(\omega)  \rangle 
&=& -2\langle I_{sd}(\omega )\rangle 
-\dfrac{\langle \delta n_s(\omega)\rangle }{\tau_{s,\mathrm{F}}} 
-\dfrac{j_{s}^\mathrm{int} (\omega) }{d_\mathrm{F}}  .
\end{eqnarray}

\noindent Using these equations the interfacial spin current can be expressed in terms of the Fourier transform of the total spin density 

\begin{equation}
\label{eq:jsi3}
j_s^\mathrm{int}(\omega)  = -2d_\mathrm{F}\widetilde{B}(\omega) 
\times 
(i\omega \langle n_{\mathrm{tot}}(\omega)\rangle ), 
\end{equation}

\noindent where the new function $\widetilde{B}(\omega) $ is given by 

\begin{equation}
\widetilde{B}(\omega) = \dfrac{B(\omega)(i\omega \tau_{s,F}+1)}
{i \omega \tau_{s,F} B(\omega) + 1} .
\end{equation}

\noindent The function $\widetilde{B}(\omega)$ carries all information about the relation between the temporal evolution of the magnetization and the interfacial spin current. We investigate the Taylor expansion

\begin{equation}
\widetilde{B}(\omega) =  B(0)+ i\tau \omega +
\mathcal{O}(\omega^2 ),
\end{equation}

\noindent where we introduced the time scale $\tau =-i\widetilde{B}'(0)$. We focus on a Ni($3.4\mbox{ nm}$)/Cu($2.5\mbox{ nm}$) bilayer, which is similar to the system used in the experiments of Ref.\  \cite{Lichtenberg2021}.  

When the interface at $x=d_\mathrm{N}$ is permeable for spins ($g_2\neq 0$) and connected to an ideal spin sink, we estimate $\tau_{g_2\neq 0}   \sim 5.1\times 10^{-16} \mbox{ s}$, when using the constants for Ni/Cu as presented in Table \ref{tab:2} and \ref{tab:3}. For frequencies up to the THz regime it satisfies $\tau_{g_2\neq 0} \,\omega\ll B(0)\sim 0.52 $, implying that $\widetilde{B}(\omega)$ is approximately independent of frequency and given by $B(0)$. Inverse Fourier transforming Eq.\ (\ref{eq:jsi3}) yields

\begin{equation}
j_s^\mathrm{int} (t) =
-2 d_\mathrm{F} B(0) \times  \dfrac{d\langle n_\mathrm{tot}\rangle }{dt} . 
\end{equation} 

\noindent By definition $-(1/s) d\langle n_\mathrm{tot}\rangle/dt  = dm/dt $. Using this substitution the interfacial spin current in terms of the normalized magnetization $m$  is

\begin{equation}
\label{eq:jsi4}
j_s^\mathrm{int} (t) = 
\epsilon\times  (2 d_\mathrm{F} s)  \dfrac{dm}{dt} ,
\end{equation}

\noindent where we defined the efficiency parameter  $\epsilon = B(0)$. This expression is identical to the relation as reported in Ref. \cite{Lichtenberg2021}.

Contrasting behavior is found when we switch to $g_2=0$, when all spins are blocked at $x=d_\mathrm{N}$. A critical role is played by the function $G_\mathrm{N}(\omega)$, which under these conditions shows $G_\mathrm{N}(0)\gg 1$ and dominates the frequency dependence of $\widetilde{B}(\omega)$. Using that the Cu nonmagnetic layer satisfies $d_\mathrm{N}/l_{s,\mathrm{N}}\ll 1$, it follows that 

\begin{equation}
\tau_{g_2=0}  \approx B(0)^2 \dfrac{\tilde{\nu}_\mathrm{F} d_\mathrm{F} \tau_{s,\mathrm{N}}^2}
   { \tilde{\nu}_\mathrm{N} d_\mathrm{N} \tau_{s,\mathrm{F}} }
   \sim 
7.0   \mbox{ fs} .
\end{equation}

\noindent In combination with $B(0)\sim 4\times 10^{-4}$, it typically satisfies $\tau_{g_2=0}\,\omega\gg B(0)$. Hence, in this specific case the first-order term of $\widetilde{B}(\omega)$ dominates. The spin current is now given by 

\begin{equation}
\label{eq:jsi5}
j_s^\mathrm{int} =  \tau_{g_2=0} \, \times  (2 d_\mathrm{F} s) 
\dfrac{d^2 m }{dt^2} .
\end{equation}

\noindent Rather than being proportional to $dm/dt$, the interfacial spin current is now approximately proportional to the second derivative of $m$. This behavior is a direct consequence of the large spin-flip scattering time $\tau_{s,\mathrm{N}}=17\mbox{ ps}$ of Cu \cite{Shin2018}. In this case an efficient spin dissipation channel is absent, resulting in an altered response of the spin accumulation in the nonmagnetic layer which affects the temporal behavior of the interfacial spin transport.   

The clear distinction between the dynamics predicted by Eq.\ (\ref{eq:jsi4}) and Eq.\ (\ref{eq:jsi5}) for a Ni($3.4\mbox{ nm}$)/Cu($2.5\mbox{ nm}$) bilayer is depicted in Figs. \ref{fig:3}(a)-(b). We here assumed that the laser pulse is only absorbed by the Ni layer and used $P_0=0.2\times 10^8 \mbox{ Jm}^{-3}$. The absorption by the Cu layer is neglected, since Cu has a relatively small imaginary component of the dielectric constant \cite{Kimling2017}. All remaining parameters are given in Table \ref{tab:2} and \ref{tab:3}. Figure  \ref{fig:3}(a) shows the correspondence between the spin current (blue solid line) and the temporal derivative of the magnetization (red dashed line) in the case that the bilayer is connected to an ideal spin sink.  There is no significant phase shift present, which is in agreement with the experiments in Ref.\ \cite{Lichtenberg2021}. In Fig.\ \ref{fig:3}(b) the spin sink is absent, and a close relation between the spin current (blue solid line) and the second derivative of the magnetization (red dashed line) is found. Note that the scaling factors  given in the figure do not match the values calculated in the text, as the calculations presented in the figures include magnon transport, the spin-dependent Seebeck effect and the full frequency dependence. 

We observe that in case the receiving layer is an efficient spin sink, or is connected to an efficient spin sink, the interfacial spin current is directly proportional to the temporal derivative of $m$, as described by the relation Eq.\ (\ref{eq:jsi4}). This is in agreement with the results of the previous section because Pt has a very short spin-flip scattering time of $\tau_{s,\mathrm{N}}\sim 0.02\mbox{ ps}$ \cite{Dang2020,freeman2018evidence}. In the opposite case, if the receiving material displays inefficient spin-flip scattering, other relations may arise. We stress that the behavior predicted by Eq.\ (\ref{eq:jsi5}) strongly depends on the exact components of the heterostructure. As shown by the calculation, a Ni/Cu bilayer is an ideal system to demonstrate the latter limiting case, mainly because Cu has a very large spin-flip scattering time scale and a Ni/Cu interface has a relatively large electrical conductance \cite{Shin2018}.  To demonstrate this experimentally, two methods can potentially be used to probe the spin-current generation into the nonmagnetic layer. First, probing the THz electromagnetic radiation that results from the inverse spin Hall effect yields the temporal profile of the spin current \cite{Rouzegar2021}. However, due to the small spin Hall angle of Cu the signal is expected to be very small \cite{wang2014scaling}. A second method is using the magneto-optical Kerr effect. In that case, the spin accumulation is probed instead of the spin current \cite{Choi2014}. For insulating boundary conditions, the spin density that builds up in the nonmagnetic layer is given by

\begin{equation}
\langle \delta n_s(\omega)  \rangle_\mathrm{N} = 
\dfrac{1}{d_\mathrm{N}} 
\dfrac{j_s^\mathrm{int} (\omega)}{ i\omega +\tau_{s,\mathrm{N}}^{-1}  }. 
\end{equation}

\noindent Here, the brackets indicate spatial averaging over the nonmagnetic layer. For Cu, with the large $\tau_{s,\mathrm{N}}$, the interfacial spin current and the build-up spin density differ a factor $\sim i\omega$. Indicating that the optically probed signal will replicate the first derivative of the magnetization. Despite the difficulty of observing the behavior of Eq.\ (\ref{eq:jsi5}), the analysis emphasizes that by modifying the properties of the nonmagnetic material the bandwidth of the spin current can be tuned \cite{Kampfrath2013,Dang2020}. Although compositions other than Ni/Cu might not yield the ideal comparison as in Figs. \ref{fig:3}(a)-(b), performing experiments for various nonmagnetic materials and probing both the magnetization and the spin current simultaneously, will yield valuable information. 

In the analytical calculation presented in this section we left out the interfacial electron-magnon scattering. In the following section, we specfically address its contribution to spin current injection and discuss the role of magnon transport.

\begin{figure*}[t!]
\includegraphics[scale=0.95]{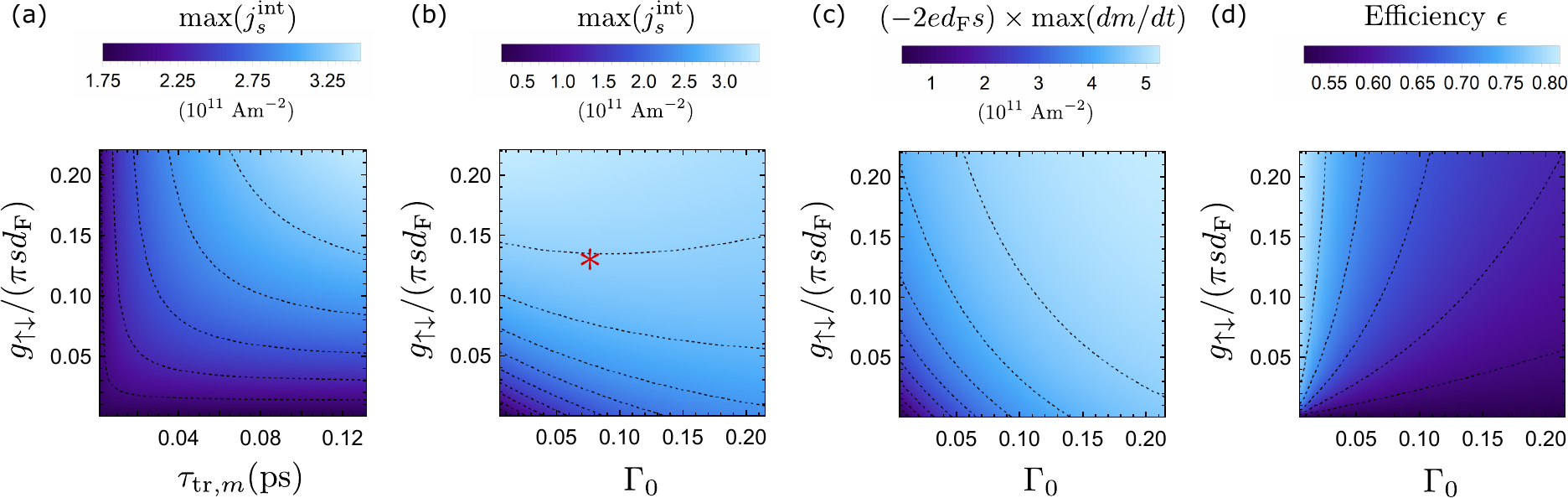}
\caption{\label{fig:4} Phase diagrams that characterize the interfacial spin current in a Ni($3.4\mbox{ nm}$)/Cu($2.5\mbox{ nm}$) bilayer connected to an ideal spin sink. (a) Phase diagram of the amplitude of the interfacial spin current as a function of the spin-mixing conductance $g_{\uparrow\downarrow} $ (made dimensionless by dividing it by $ (\pi s d_\mathrm{F}) $)  and the magnon transport time scale $\tau_{\mathrm{tr},m}$.  (b)-(d) Phase diagrams as a function of the spin-mixing conductance $g_{\uparrow\downarrow}$ and the bulk electron-magnon scattering rate coefficient $\Gamma_0$, where $\tau_{\mathrm{tr},m}=0.1\mbox{ ps}$ is set constant. (b) The amplitude of the interfacial spin current. The red star indicates the values used in Fig.\ \ref{fig:3}(a). (c) The amplitude of $(-2 e d_\mathrm{F} s) dm/dt$. (d) The efficiency parameter $\epsilon$, as defined in Eq.\ (\ref{eq:jsi4}).  
} 
\end{figure*}


\subsubsection{Interfacial electron-magnon scattering and magnon transport} 

\noindent Magnon spin transport and magnon heat transport are directly coupled, which makes them complex to investigate analytically. Hence, this section presents a  numerical analysis of the role of interfacial electron-magnon scattering and magnon transport. 

The results are shown in Figs. \ref{fig:4}(a)-(d), and correspond to a Ni($3.4\mbox{ nm}$)/Cu($2.5\mbox{ nm}$) bilayer connected to an ideal spin sink. The  calculations include bulk electron-magnon scattering, interfacial electron-magnon scattering and the spin-dependent Seebeck effect. The used system parameters are presented in Table \ref{tab:2} and \ref{tab:3}. The phase diagram in Fig.\ \ref{fig:4}(a) indicates the amplitude of the spin current, determined by its maximum value, as a function of the spin-mixing conductance  $g_{\uparrow\downarrow}$ and the magnon transport time scale $\tau_{\mathrm{tr},m}$. The spin-mixing conductance, which is made dimensionless by dividing it by a factor $(\pi s d_\mathrm{F})$, determines the strength of the interfacial electron-magnon scattering and thereby the magnon spin current near the interface. The time scale $\tau_{\mathrm{tr},m}$ determines the effectiveness of magnon transport in the bulk. In order to have interfacial electron-magnon scattering significantly contribute to the total spin current, it is required to have  efficient magnon transport in the bulk, as indicated by the light region in Fig.\ \ref{fig:4}(a). 

Figure \ref{fig:4}(b)-(d) compare the contributions by bulk and interfacial electron-magnon scattering. These phase diagrams are plotted as a function of the spin-mixing conductance and the bulk electron-magnon scattering rate coefficient $\Gamma_0$. Here, the magnon transport time scale is set to $\tau_{\mathrm{tr},m}=0.1\mbox{ ps}$, similar to the calculations presented in the previous sections. The color scheme in Fig.\ \ref{fig:4}(b) indicates the amplitude of the interfacial spin current. The figure   emphasizes that including interfacial electron-magnon scattering boosts the amplitude of the spin current, however, the significance of the increase depends on the efficiency of the bulk electron-magnon scattering. 

Figure \ref{fig:4}(c) indicates the amplitude of $dm/dt$, given in the same units as the spin current. Intuitively, the demagnetization favours a simultaneously large interfacial and bulk electron-magnon scattering rate, since both contribute to the generation of thermal magnons. This does not linearly translate to a maximized spin current, as the relation between the spin current and demagnetization depends on which contribution dominates. The color scheme in Fig.\ \ref{fig:4}(d) indicates the efficiency $\epsilon$, as defined in Eq.\ (\ref{eq:jsi4}). Keeping in mind the analytical calculation, the range of $\epsilon$ is approximately related to the values of prefactors $A(0)$ and $B(0)$ (see Eqs.\ (\ref{eq:A})-(\ref{eq:B})). In case only interfacial electron-magnon scattering is present this yields $\epsilon\sim A(0)\sim 0.79 $, whereas for the pure bulk scenario $\epsilon\sim B(0)\sim 0.52$. A small deviation compared to Fig.\ \ref{fig:4}(d) arises as the numerical calculation includes the spin-dependent Seebeck effect and the full frequency dependence. 

All calculations presented here imply that spin transport by magnons, which is typically neglected in the calculations of laser-induced spin transport in metallic magnetic heterostructures \cite{Shin2018,Kimling2017}, is relevant to include in the analyses \cite{Tveten2015}. Since magnon transport is driven by electron-magnon scattering at the interface, the ratio of $\Gamma_0$ and $g_{\uparrow\downarrow}/(\pi s d_\mathrm{F})$ plays a decissive role \cite{Tveten2015}. Furthermore, constants that parametrize either bulk magnon transport or spin-polarized electron transport are essential. Their coupled dynamics complexifies the characterization of bulk spin transport, including the modification of the diffusion length scales \cite{cheng2017interplay}. On top of that, nonmagnetic system parameters that correspond to the  thermal proporties of the system do strongly affect the importance of magnon transport. For instance, the interfacial magnon current is mainly determined by the temperature difference $T_m-T_e^\mathrm{N}$, which critically depends on the thermal and optical proporties of the nonmagnetic layer. Further theoretical work is required to chart the essential dependencies on the  properties of the heterostructure. 
 
Hence, we state that the role of interfacial electron-magnon scattering and, consequently, bulk magnon transport can not be neglected a priori \cite{Tveten2015}. It may play a significant role dependent on the specific system components and properties. Nevertheless, bulk electron-magnon scattering always remains essential, as ultrafast demagnetization is observed in magnetic thin films regardless of the presence of a neighbouring nonmagnetic metallic layer \cite{Schellekens2013superdiffusive}.

\section{Conclusion and outlook } 

In conclusion, we modeled ultrafast demagnetization and spin transport in rapidly heated magnetic heterostructures, addressing the interplay of thermal magnons and itinerant spins. Within this model, the magnetization is determined by the total spin density of the two populations and ultrafast demagnetization is driven by the combination of electron-magnon scattering and additional spin-flip scattering processes originating from spin-orbit coupling. Secondly, electron-magnon scattering is a driving force of nonlocal spin transfer, for which we calculated the resulting spin transport by magnons and spin-polarized electrons within a diffusive description. It is shown that, in case the receiving material is an efficient spin sink, the interfacial spin current becomes directly proportional to the temporal derivative of the magnetization. Furthermore, we have discussed the role of interfacial electron-magnon scattering and magnon transport, and showed that they cannot be neglected a priori. However, their significance strongly depends on the material properties of the full magnetic heterostructure. 

In this work we focused on ultrathin magnetic heterostructures. To explore the role of bulk temperature gradients and identify  characteristic length scales, a quantitative analysis over a larger range of thicknesses is required. Secondly, it will become interesting to go beyond the assumptions that the phononic system plays a minor role and behaves as an ideal spin sink. As recent experiments show that during the ultrafast demagnetization spin is transferred to the lattice \cite{Dornes2019,Tauchert2021}, and specifically circularly polarized phonons \cite{Tauchert2021}, it becomes obvious that a more complete description of the phononic system is needed. Moreover, it was already proposed that a coupling between magnons and phonons should be inplemented within a three-temperature description \cite{Kang2020}. Nevertheless, it is expected that the dominant physical concepts are captured within the assumptions of the presented model. 

This work is part of the research programme of the Foundation for Fundamental Research on Matter (FOM), which is part of the Netherlands Organisation for Scientific Research (NWO). R.D. is member of the D-ITP consortium, a program of the NWO that is funded by the Dutch Ministry of Education, Culture and Science (OCW). This work is funded by the European Research Council (ERC).

\bibliographystyle{apsrev4-2}

\providecommand{\noopsort}[1]{}\providecommand{\singleletter}[1]{#1}%

\appendix
\setcounter{table}{0}
\renewcommand{\thetable}{A\arabic{table}}

\section{Expansion of the polylogarithm}
\label{sec:appA}

\begin{table}[b!]
\centering
\caption{\label{tab:2} Parameters that characterize the magnonic system in Ni.}
\begin{tabular}[t]{lcc}
\hline \\ [-2.0ex] 
symbol &  meaning & estimate  \\
\hline \\ [-1.5ex] 
$T_0\,(\mbox{K})$  & ambient temperature & $295$ \\
$T_C\,(\mbox{K}) $ \cite{Coey2009} & Curie temperature  & $628$  \\ 
$\tau_{\mathrm{tr},m}\,(\mbox{ps}) $ \footnotemark[1]  & magnon momentum relaxation time  & $0.1$ \\ 
$A\,(\mbox{meV\AA}^{2}) $ \footnotemark[2]  & spin-wave stiffness & $400 $  \\
$a\,(\mbox{nm}) $  \cite{davey1925precision}  & lattice constant & $0.35$ \\
$S$ \footnotemark[3] & Spin per atom (in units $\hbar$) & $0.6\times (1/2)$  \\
$\Gamma_0$ \footnotemark[4] & e-m scattering rate coefficient & $2\times 0.038$  \\
$\epsilon_0\,(\mbox{meV}) $ \footnotemark[5]  & magnon gap& $0.05$ \\
\hline
\footnotetext{Discussed in the main text.} 
\footnotetext{Typical order of magnitude estimated by $A\sim 2 k_B T_C S a^2$.}
\footnotetext{Estimated from atomic magnetic moment given in \cite{Koopmans2010}.}
\footnotetext{Using relation  $\Gamma_0=2\alpha$ \cite{Bender2014}  and $\alpha$ of Ni \cite{Koopmans2005}.} 
\footnotetext{Typical order of magnitude from  FMR frequency of $\sim 10 \mbox{ GHz}$.}  
\end{tabular}
\end{table}%

\noindent To evaluate the integrals that are needed for the description of the magnonic system, we make use of the following expression for the polylogarithm \cite{Bender2012}

\begin{equation}
\mbox{Li}_s(e^x) 
= \dfrac{1}{\Gamma(s)} 
\int_0^\infty 
dt 
\dfrac{t^{s-1}}{e^t/e^x-1} .
\end{equation}

\noindent As mentioned in Sec.\ \ref{sec:2a}, we assume the parameter $x$ remains small. The polylogarithm can be written as a series expansion \cite{basso2016thermodynamic}

\begin{equation}
\label{eq:a2}
\mbox{Li}_s(e^x) =
\Gamma(1-s) (-x)^{s-1} + 
\sum_{k=0}^\infty 
\dfrac{\zeta(s-k)}{k!}
x^k .
\end{equation}

\noindent Since $x$ remains small, we truncate this series for $k\geq 2$, which is the basis for the calculation of all coefficients presented in Table \ref{tab:1}. As an example, we calculate the magnon density from Eq.\ (\ref{eq:nd0}) 

\begin{eqnarray}
\label{eq:a4}
n_d = \dfrac{(k_B T_m)^{3/2}}{4\pi^2 A^{3/2}} 
\Gamma(3/2) 
\mbox{Li}_{3/2} \bigg(
e^{(\mu-\epsilon_0)/(k_B T_m)} 
\bigg).
\end{eqnarray}

\noindent Applying the expansion up to first order in $\mu_m/(k_B T_m),\epsilon_0/(k_B T_m) \ll 1 $ we have

\begin{eqnarray}
n_d &=& \dfrac{(k_B T_m)^{3/2}}{4\pi^2 A^{3/2}} 
\Gamma(3/2) 
\bigg[\Gamma(-1/2) \bigg(
\dfrac{\epsilon_0-\mu_m}{k_B T_m} 
\bigg)^{1/2} 
\\
\nonumber
&& \qquad \qquad \qquad  \qquad 
+\zeta(3/2)-\zeta(1/2) \bigg(
\dfrac{\epsilon_0-\mu_m}{k_B T_m} 
\bigg)
\bigg] .
\end{eqnarray}

\noindent If we now impose that we only have small changes of the magnon temperature compared to room temperature, $(T_m-T_0)/T_0\ll 1 $, and only collect the terms up to first order in small parameters we find

\begin{eqnarray}
\label{eq:a5}
n_d &=& \dfrac{(k_B T_0)^{3/2}}{4\pi^2 A^{3/2}} 
\Gamma(3/2) 
\\
\nonumber 
&&
\bigg[\Gamma(-1/2) \bigg(
\dfrac{\epsilon_0-\mu_m}{k_B T_0} 
\bigg)^{1/2} 
+\zeta(3/2) 
\\
\nonumber 
&&- 
\zeta(1/2) \bigg(
\dfrac{\epsilon_0-\mu_m}{k_B T_0} 
\bigg)
+(3/2) \zeta(3/2) \dfrac{(T_m-T_0)}{ T_0}
\bigg]. 
\end{eqnarray}  

\noindent Evaluatimg this expression at the maximum temperature and chemical potential of the calculation in Sec.\ \ref{sec:3a} (Fig.\ \ref{fig:1}), it only differs approximately one per cent from the exact value Eq.\ (\ref{eq:a4}). 

Taking the temporal derivative of $n_d$ yields

\begin{eqnarray}
\dfrac{\partial n_d}{\partial t} &=& \dfrac{(k_B T_0)^{3/2}}{4\pi^2 A^{3/2}} 
\Gamma(3/2) 
\\
\nonumber 
&&
\bigg[\bigg(\Gamma(1/2) \bigg(
\dfrac{\epsilon_0-\mu_m}{k_B T_0} 
\bigg)^{-1/2} 
+\zeta(1/2) \bigg)\dfrac{\dot{\mu}_m}{k_B T_0}  
\\
\nonumber 
&&\qquad 
+(3/2) \zeta(3/2)\dfrac{\dot{T}_m}{ T_0}
\bigg],
\end{eqnarray}

\noindent which determines the coefficients $C_{n,\mu}$ and $C_{n,T}$, as defined in the main text and given in Table \ref{tab:1}.  When $\mu_m$ approaches the magnon gap $\epsilon_0$, the first term in $C_{n,\mu}$ diverges, which originates from Bose-Einstein statistics. It is essential to include this nonlinear term in $C_{n,\mu}$ as otherwise we would find time traces of the magnon chemical potential that may largely exceed the magnon gap.

For all remaining coefficients in Table \ref{tab:1}, the first term in  the expansion Eq.\ (\ref{eq:a2}) will only yield higher order contributions. In that case the coefficients follow equivalently from first-order Taylor expansion (the second term in Eq.\ (\ref{eq:a2})), where in the prefactors it is used that for sufficiently small $\epsilon_0$ we approximate $\mbox{Li}_s(\exp(-\epsilon_0/(k_B T_0))) \sim \zeta(s) $. We stress that all expansion methods we use here remain a rough estimate. In order to retrieve valuable quantitative results from the magnonic calculation it is essential to implement the full polylogarithm.

\section{Notes on the approximations in the analytical calculation } 
\label{sec:appB}

\noindent In Section \ref{sec:3c}  the bulk  electron-magnon scattering rate, which is implemented in the spin diffusion equation as a source of spins, is simplified using the following considerations. We express the source $I_{sd} (\omega,x) $ in terms of a cosine expansion 

\begin{eqnarray}
I_{sd}(\omega,x) &=& 
\dfrac{I_{sd,0}(\omega) }{2} 
+ \sum_{n=1}^\infty 
I_{sd,n}(\omega) 
\cos\Big(\dfrac{n \pi x}{d_\mathrm{F}} \Big) ,
\end{eqnarray}

\noindent where the coefficients $I_{sd,n} (\omega) $ are given by

\begin{eqnarray}
I_{sd,n} &=& \dfrac{2}{d_\mathrm{F}} 
\int_{-d_\mathrm{F}}^0 dx I_{sd}(\omega,x) \cos\Big(\dfrac{n \pi x}{d_\mathrm{F}} \Big).
\end{eqnarray}

\noindent Note that the zeroth mode corresponds to twice the spatial average $I_{sd,0}(\omega)= 2\langle I_{sd}(\omega)\rangle $. The higher-order modes are a measure of the spatial inhomogeneity of the source term. We want to express the function $\mathcal{I}_{sd} (\omega)$, as given in Eq.\ (\ref{eq:source}), in terms of the coefficients $I_{sd,n}(\omega)$. By performing the spatial integration we find 

\begin{eqnarray}
\label{eq:b3}
\mathcal{I}_{sd}(\omega) 
&=& \langle I_{sd} (\omega) \rangle 
+ 
\sum_{n=1}^\infty 
\dfrac{I_{sd,n} (\omega) }{\Big( \dfrac{n \pi }{d_\mathrm{F}\kappa_\mathrm{F}(\omega) } \Big)^2 +1}. 
\end{eqnarray}

\noindent Hence, the $n\geq 1$ modes of $I_{sd}(\omega,x)$ are truncated by the denominator. In combination with that the inhomogeneous modes remain relatively small compared to the homogeneous mode, it turns out to be a relevant approximation to neglect all the terms in the summation in Eq.\ (\ref{eq:b3}).


\section{System parameters}
\label{sec:appC}

\noindent The system parameters that are used in the calculations presented in the main text are summarized in Table \ref{tab:2} and Table \ref{tab:3}. Table \ref{tab:2} shows the estimated parameters that characterize the magnonic system in Ni. Table \ref{tab:3} presents the parameters of the electronic system in Ni, Pt and Cu. Furthermore, it includes the parameters that correspond to the interfaces.

\begin{table}[t!]
\centering
\caption{\label{tab:3} Parameters for the electronic system of Ni, Pt and Cu. Parameters that characterize the interface correspond to Ni/Cu and Ni/Pt.}
\begin{tabular}[t]{lcccc}
\hline \\ [-2.0ex] 
symbol & Ni & Pt & Cu & Ref.\ \\ 
\hline \\ [-1.5ex] 
$\gamma\,(\mbox{Jm}^{-3}\mbox{K}^{-2})$ &$ 1077 $ & $721$ & $100$ & \cite{Shin2018,Kang2020,Kimling2017} \\ 
$C_p\,(10^6\,\mbox{Jm}^{-3}\mbox{K}^{-1})$ &  $ 3.6  $ & $2.85 $  &$3.45   $ & \cite{Kang2020,Kimling2017} \\ 
$g_{ep}\, (10^6\,\mbox{Jm}^{-3}\mbox{ps}^{-1})$ & $ 0.855  $ & $ 0.29  $ &  $ 0.07  $ & \cite{Kang2020,Kimling2017} \\ 
$2\tilde{\nu}\,(\mbox{eV}^{-1}\mbox{nm}^{-3}) $ \footnotemark[1] & $ 272 $ &$ 137 $\footnotemark[2] & $  26 $ & \cite{Dang2020,ko2020optical,Shin2018} \\
$2\tilde{\sigma} \,(10^6\,\mbox{Sm}^{-1}) $ \footnotemark[1] & $  7.1$ &$  6.6 $ & $  39 $ & \cite{Shin2018,Kimling2017} \\
$\kappa_e \,( \mbox{Wm}^{-1}\mbox{K}^{-1} ) $ & $ 50 $ &$   50 $ & $  300 $ & \cite{Shin2018,Kimling2017} \\
$S_s \,(10^{-24}\, \mbox{JK}^{-1}) $ \footnotemark[3] & $ -0.3 $ & &  &  \\
$g_1 \, (10^{19}\,\mbox{m}^{-2})  $ \footnotemark[4] &  & $0.3  $ & $1.0  $& \cite{Shin2018,Kimling2017}\\
$g_2 \,(10^{19} \,\mbox{m}^{-2} )$ &    &   &  $1.0  $ & \\
$\kappa_e^i \,(10^9\,\mbox{Wm}^{-2}\mbox{K}^{-1}) $ \footnotemark[5] &   & $10$ & $40 $& \cite{Kimling2017,Shin2018} \\
$S^i_s\,(10^{-24}\,\mbox{JK}^{-1} )$ \footnotemark[6] &   &  $-0.3 $ &  $-0.3$  & \\
$g_{\uparrow\downarrow}\,(10^{19}\,\mbox{m}^{-2}) $   &  & $0.3   $ & $1.0  $& \cite{tserkovnyak2005nonlocal,yoshino2011quantifying} \\
$\tau_s\, (\mbox{ps})  $   & $0.1$ & $0.02 $ & $17 $& \cite{Shin2018,Dang2020,freeman2018evidence} \\
\hline
\footnotetext{ We assume that the spin-averaged quantities $\tilde{\nu}$ and $\tilde{\sigma}$ are approximately given by the (total) electrical quantity divided by two. } 
\footnotetext{Calculated from the ratio of the conductivity and diffusion coefficient in \cite{Dang2020}. }
\footnotetext{Using that the Seebeck coefficient scales as $(\pi^2/3)k_B (T_0/T_F)$, with Fermi temperature $T_F\sim 10^4\mbox{ K}$, the polarization $P_s\sim 0.2$ and sign of the spin-dependent Seebeck effect \cite{Slachter2010}. }
\footnotetext{Estimated from the electrical conductance given for Ni/Cu in \cite{Shin2018} and [Co/Ni]/Pt in \cite{Kimling2017}. }
\footnotetext{Estimated from the electrical conductance and the Wiedemann-Franz law \cite{Shin2018}. } 
\footnotetext{Assumed to be equal to the bulk value. } 
\end{tabular}
\end{table}%

\end{document}